\documentclass[preprint,amsmath,amssymb,superscriptaddress,amsfonts,nofootinbib,showpacs,prc,floatfix]{revtex4-1}

\usepackage{graphicx}
\usepackage{dcolumn}
\usepackage{bm}

\begin{document}


\title{Dipolar degrees of freedom and Isospin equilibration processes in Heavy Ion collisions.}
\author{M.Papa}\email{papa@ct.infn.it}
\affiliation{INFN, Catania, Italy}
\author{I.~Berceanu}
\affiliation{National Institute for Physics, Bucharest, Romania}
\author{L.~Acosta}
\affiliation{INFN-Laboratori Nazionali del Sud, Catania, Italy}
\author{F.~Amorini}
\affiliation{INFN-Laboratori Nazionali del Sud, Catania, Italy}
\author{C.~Agodi}
\affiliation{INFN-Laboratori Nazionali del Sud, Catania, Italy}
\author{A.~Anzalone}
\affiliation{INFN-Laboratori Nazionali del Sud, Catania, Italy}
\author{L.~Auditore}
\affiliation{INFN, Messina, Italy}
\affiliation{Universit$\grave{a}$ facolt$\grave{a}$ di Fisica, Messina,Italy}
\author{G.~Cardella}
\affiliation{INFN, Catania, Italy}
\author{S.~Cavallaro}
\affiliation{INFN-Laboratori Nazionali del Sud, Catania, Italy}
\author{M.B.~Chatterjee}
\affiliation{Saha, Institute of Nuclear Physics Kolkata, India}
\author{E.~De~Filippo}
\affiliation{INFN, Catania, Italy}
\author{L.~Francalanza}
\affiliation{INFN-Laboratori Nazionali del Sud, Catania, Italy}
\affiliation{Universit$\grave{a}$ facolt$\grave{a}$ di Fisica e Astronomia, Catania, Italy}
\author{E.~Geraci}
\affiliation{INFN, Catania, Italy}
\affiliation{Univerist$\grave{a}$ facolt$\grave{a}$ di Fisica e Astronomia, Catania, Italy}
\author{L.~Grassi}
\affiliation{INFN, Catania, Italy}
\affiliation{Universit$\grave{a}$ facolt$\grave{a}$ di Fisica e Astronomia, Catania, Italy}
\author{B.~Gnoffo}
\affiliation{INFN, Catania, Italy}
\affiliation{Universit$\grave{a}$ facolt$\grave{a}$ di Fisica e Astronomia, Catania, Italy}
\author{J.~Han}
\affiliation{INFN-Laboratori Nazionali del Sud, Catania, Italy}
\author{E.~La~Guidara}\affiliation{INFN, Catania, Italy}
\author{G.~Lanzalone}
\affiliation{INFN-Laboratori Nazionali del Sud, Catania, Italy}
\affiliation{``Kore'' Universit$\grave{a}$, Enna, Italy}
\author{I. Lombardo}
\affiliation{Universit$\grave{a}$ facolt$\grave{a}$ di Fisica, Napoli, Italy}
\author{C.~Maiolino}
\affiliation{INFN-Laboratori Nazionali del Sud, Catania, Italy}
\author{T.~Minniti}
\affiliation{INFN, Messina, Italy}
\author{A.~Pagano}
\affiliation{INFN, Catania, Italy}
\author{E.V.~Pagano}
\affiliation{INFN-Laboratori Nazionali del Sud, Catania, Italy}
\affiliation{Universit$\grave{a}$ facolt$\grave{a}$ di Fisica e Astronomia, Catania, Italy}
\author{S.~Pirrone}
\affiliation{INFN, Catania, Italy}
\author{G.~Politi}
\affiliation{Universit$\grave{a}$ facolt$\grave{a}$ di Fisica e Astronomia, Catania, Italy}
\author{F.~Porto}
\affiliation{INFN-Laboratori Nazionali del Sud, Catania, Italy}
\affiliation{Universit$\grave{a}$ facolt$\grave{a}$ di Fisica e Astronomia, Catania, Italy}
\author{L.~Quattrocchi}
\affiliation{INFN, Messina,  Italy}
\author{F.~Rizzo}
\affiliation{INFN-Laboratori Nazionali del Sud, Catania, Italy}
\affiliation{Universit$\grave{a}$ facolt$\grave{a}$ di Fisica e Astronomia, Catania, Italy}
\author{E. Rosato}
\affiliation{Universita' facolta di Fisica, Napoli, Italy}
\affiliation{Universit$\grave{a}$ facolt$\grave{a}$ di Fisica, Napoli, Italy}
\author{P.~Russotto}
\affiliation{INFN-Laboratori Nazionali del Sud, Catania, Italy}
\author{A.~Trifir\`o}
\affiliation{INFN, Messina, Italy}
\affiliation{Universit$\grave{a}$ facolt$\grave{a}$ di Fisica, Messina,Italy}
\author{M.~Trimarchi}
\affiliation{Universit$\grave{a}$ facolt$\grave{a}$ di Fisica, Messina,Italy}
\affiliation{Universita' facolta di Fisica, Napoli, Italy}
\author{G.~Verde}
\affiliation{INFN, Catania, Italy}
\author{M.~Vigilante}
\affiliation{INFN, Napoli, Italy}
\date{\today}

%
%



\begin{abstract}
\begin{description}
\item[Background] In heavy ion collision at the Fermi energies Isospin equilibration processes
 occurring when nuclei with different charge/mass asymmetries interacts have been investigated to get information
 on the nucleon-nucleon Iso-vectorial effective interaction.

\item[Purpose] In this paper, for the system $^{48}Ca +^{27}Al$ at 40 MeV/nucleon, we investigate on this process by means of an observable tightly linked to
 isospin equilibration processes  and sensitive in exclusive way to the dynamical stage of the collision.
From the comparison with dynamical model calculations we want also to obtain information on the Iso-vectorial effective microscopic interaction.
\item[Method] The average time derivative of the total dipole associated to the relative motion of all emitted  charged particles and fragments has been determined from the measured charges and velocities by using the $4\pi$ multi-detector CHIMERA. The average has been determined for semi-peripheral collisions and for different charges $Z_{b}$ of the biggest produced fragment.
Experimental evidences collected for the systems
$^{27}Al+^{48}Ca$ and $^{27}Al+^{40}Ca$ at 40 MeV/nucleon used to support this novel method
of investigation are also discussed.
\item[Results] The data analysis shows a clear signature of a trend to the
 global Isospin equilibration of the system for increasing differences of the $Z_{b}$ values with respect to
  the projectile charge.
\item[Conclusions]
The comparison with CoMD-II calculations gives the best agreement with data using a  stiffness $\gamma$ value for the Iso-vectorial interaction in the range $\gamma\simeq 1-1.2$.
 Moreover, the same comparison allows to estimate the non-negligible contribution to the global isospin equilibration process given by the un-detected
  emitted neutrons.
\end{description}
\end{abstract}

\maketitle


\section{Introduction}
Experimental evidences on Heavy Ion Collisions highlight, in different aspects,
processes evolving on different time scales.
At the Fermi energies  semiclassical dynamical models can not
describe the system during its overall time evolution.
According to the displayed phenomenology, the collected data  are usually described as produced by a fast pre-equilibrium stage
described through dynamical models \cite{baorep,barrep,bnv} and later stage processes described by statistical decay models \cite{smm}. The statistical contribution is usually separated from the dynamical one through cut/extrapolation procedures applied to
angular correlation  and/or to particle kinetic energy spectra (see as an example \cite{piant,russ}).
When clearly identified \cite{g1}, the attempt to measure observables in principle closely linked to only one
of the two regime is therefore highly desirable. In fact, this eventuality allows
to decouple effects related to the two classes of mechanisms that are linked to
rather different nuclear matter macroscopic properties.
In the last decades great efforts have been performed to extract information on the nuclear iso-vectorial
forces by studying  charge/mass asymmetric systems
\cite{dan,admard1,admard2,tsang,marini,kohley,lim1,lim2,natw,yen,levd1}.
These attempts concerns  both  the dynamical stage and the statistical decay of the produced
hot sources. In particular in this last stage  the isospin and excitation energy dependence of the level density formula play a key role and it is currently  under investigation \cite{admard2,marini,levd1,napoli,isodec}.

 A phenomenon closely linked with iso-vectorial-forces is the well known process leading to the redistribution
 in phase-space of the charge/mass excess $\beta=\frac{N-Z}{A}$ \cite{rami}  of the emitted particles and fragments
 (Z is the charge, N is the neutron number and A the mass number).
This phenomenon commonly referred as
 charge/mass or isospin equilibration process is rather complex in the Fermi energy domain, especially when finite
 size effects of the studied system have to be properly taken into account.
  The  charge/mass distributions related to the final fragments and particles are affected by the pre-equilibrium stage, which includes
particles and fragments production in the mid-rapidity region, prompt emission, transfer of mass and charge between the main fragments and finally
particles/fragments emission from  hot equilibrated sources through a multi-step statistical cascade.
As an example, in Refs \cite{rami,tsang,natw,lomb,sun} the transfer of mass-charge between the main partners has been investigated
through the study of the isospin transport ratio obtained starting from the isotopic distributions produced near the projectile
rapidity for medium-heavy symmetric/quasy-symmetric systems.

In this paper we report on results of investigations on this equilibration process  for the system $^{48}Ca +^{27}Al$ at 40 MeV/nucleon starting from a different and/or complementary point of view.
The measurement was performed with the CHIMERA multi-detector \cite{chi1} at Laboratori Nazionali del Sud di Catania (Italy). The main goal of the experiment was to evaluate, for well reconstructed events, belonging to selected classes $\mathcal{K}$, the
quantity:
\begin{equation}\label{1}
\langle\overrightarrow{D}\rangle=\langle\sum_{i=1}^{m}Z_{i}
(\overrightarrow{V}_{i}-\overrightarrow{V}_{c.m.})\rangle_{\mathcal{K}}.
\end{equation}

The brackets indicate the average value over the ensemble $\mathcal{K}$.
$Z_{i},V_{i},m$ are the charges, laboratory velocities, charged particle multiplicity respectively of the produced
particles in the selected class of events, respectively. Finally $\overrightarrow{V}_{c.m.}$ is the center of mass (c.m.) velocity.
We note that in this expression the contribution of produced neutral particles is implicitly
contained in $\overrightarrow{V}_{c.m.}$.
The interest on this quantity was triggered by two main reasons:

a) as shown in Refs \cite{g1,rap} this quantity is closely linked with charge/mass equilibration process because it represents
the average time derivative of the total dipolar signal  in the asymptotic stage (expressed in unit of $e$). In fact, as an example, for binary
systems, in absence of dynamical neutron-proton collective motion we have:
$\langle\overrightarrow{D}\rangle \equiv \overrightarrow{D}_{m} = \frac{1}{2}\langle\mu\rangle(\langle\beta_{2}\rangle-\langle\beta_{1}\rangle)(\langle\overrightarrow{V}_{1}-\overrightarrow{V}_{2})\rangle$.
$\mu$ is the reduced mass number of the system, $\beta_{1},\beta_{2}$  are the isospin asymmetries of the two partners 1 and 2
and finally  $\overrightarrow{V}_{1}$ and $\overrightarrow{V}_{2}$  the related velocities.
In the above expression we have  supposed negligible the correlation of fluctuations between charge/mass ratios of the partners and their relative velocity.
On the other limit, the same quantity is zero if evaluated for a system represented by an equilibrated source before or
after the statistical decay.
As shown from dynamical microscopic calculations  in a collision process between two
nuclei having different charge/mass asymmetries, $\mid\langle\overrightarrow{D}\rangle\mid$  changes during the time
towards smaller values in the pre-equilibrium stage (spontaneous approach to the equilibrium)
producing $\gamma$-ray emission through the excitation of a more or less damped dipolar dynamical mode \cite{g1,g2};

b) because of the symmetries of the statistical decay mode, $\langle\overrightarrow{D}\rangle$ is not
affected by the statistical emission  of all the produced sources in later stages,
as it is shown in Ref. \cite{g1}. This essentially happens because, due to the vectorial kinematical character
of this quantity , for well reconstructed events statistical effects are self-averaged to zero.
Therefore  $\langle\overrightarrow{D}\rangle$ is a rather well suited global variable to selectively evidence
dynamical effects related to the Isospin equilibration process.

\section{ The Experimental procedure and data analysis}
The experiment was carried out by using
40 MeV/nucleon $^{48}Ca$ and $^{27}Al$ beams,
 at the LNS Super-Conducting Cyclotron. The beam impinged on 400$\frac{\mu g}{cm^{2}}$ $^{27}Al$ and about 1200$\frac{\mu g}{cm^{2}}$ $^{48}Ca$,$^{40}Ca$ targets.
 The chosen combinations were the following: $^{48}Ca+^{27}Al$ as the main system to be investigated,
 the $^{27}Al+^{48}Ca$ as the  auxiliary system (see the following)
 and the charge/mass quasi-symmetric $^{27}Al+^{40}Ca$
 as reference system.
 Charges, masses, energies and velocities of the produced particles and fragments were measured with the
 4$\pi$  Multi-Detector CHIMERA \cite{chi1,chi2}. In particular,
the $\Delta E-E$ technique was employed for Z identification
of fragments punching through the silicon detectors and
additionally for isotopic identification of fragments with atomic
numbers $Z <$ 10.
Mass identification is performed with the time-of-flight
(TOF) technique  by using the time signal from silicon
detectors with respect to the time reference of the radio-frequency
signal from the cyclotron. The TOF technique is
basically used for velocity measurements of heavy ions. This technique is also essential
for the mass and  charge indirect determination  of slow TLF's (Target Like Fragments)
stopped in the silicon detectors.
Energetic light charged particles, stopped in the
scintillator crystal, are identified by applying the "fast-slow"
discrimination method \cite{alder}.

In the following we report results collected for events with a multiplicity of detected charged particles
greater than or equal to 2.
We have chosen rather restrictive selection criteria to identify the "good" reconstructed events.
These  conditions  are suggested from calculations to obtain  the "invariance"
of the investigated quantity with respect the statistical decay mode (see Table I).
For the main system we have selected events for which the total identified charge $Z_{tot}^{d}=33$. Checks are being made to see at what extent the previous condition can be slightly released always keeping the above mentioned "invariance" condition to an acceptable level.
The total detected mass was chosen in  the interval $62\leq A_{tot}^{d}\leq 78$.
The total measured momentum along the beam axis has been selected
within 70$\%$ of the theoretical value (422 amu $\cdot$ cm/nsec).

Analogous conditions have been imposed for the others investigated systems,
taking into account the differences in the total mass and c.m. velocity.
The well reconstructed events have been classified  according to the charge of the biggest detected  fragment $Z_{b}$
and according to the estimated total kinetic energy loss
$TKEL=\mu E^{in}_{A}-\sum_{i=1}^{m}\frac{1}{2}(M_{i}V_{i}^{2}-M_{tot}^{d}V_{c.m.,d}^{2})$.
Where $E^{in}_{A}$ is the incident energy per nucleon and $\mu$ is reduced mass number of the impinging nuclei.
In the above expression $M_{i}$,$V_{i}$ are the measured masses and velocities; $M_{tot}^{d}$ is the total measured mass and $\overrightarrow{V}_{c.m.,d}$ the related c.m. velocity associated to the detected charged particles. Therefore, each event has been
characterized also through these two last quantities,
that in the TKEL evaluation can globally compensate the uncertainties
due to the non perfect  mass identification and velocity measurement.
For the main system $^{48}Ca+^{27}Al$ and the auxiliary one $^{27}Al+^{48}Ca$,
in Fig.1 we show the charge of the detected fragments as function of their velocity $V_{Z}$ along the beam direction.
In both cases the bi-dimensional plots show the dominance of  processes producing TLF's and PLF's (Projectile Like fragments).
The thin black curves represent
the threshold of the $\Delta E-E$ charge  identification technique as function of the fragments velocity evaluated
for the typical Silicon detector thickness  of the CHIMERA apparatus (about 300$\mu m$).

For each velocity value, fragments having a charge smaller than the value plotted through the curves
can be directly assigned by means of the $\Delta E-E$ technique.
Fragments having an higher charge will be stopped in the silicon detector an then the charge identification is obtained in an indirect way through the mass determination obtained by means of TOF and energy measurements and by using the Charity prescription \cite{char}.
 \begin{figure}
 \centering
\includegraphics[scale=0.43]{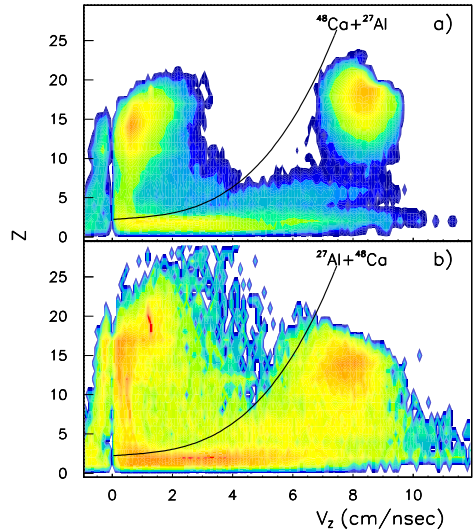}
\caption{\label{fig1} (Color online) Charges Z of the
detected fragments versus
their velocities $V_{Z}$ along the beam axis are shown
for the main system (panel a))
and the auxiliary one (panel b).}
\end{figure}

Therefore the comparison between the TLF charge distribution obtained in the main system in the indirect way (upper portion of the Z-$V_{Z}$ plot in Fig.1(a)) and the one related to the PLF  for the auxiliary system, directly assigned,
(lower portion of Z-$V_{Z}$ plot in Fig.1(b)) allows to evidence eventual systematic errors in the indirect charge assignment associated to the main system.
The comparison has been performed for different windows of TKEL.

It results that for TKEL lower than about 350 MeV the obtained charge distributions are  similar and
no  systematic error is evidenced. This result is shown in Fig.2 in panels a) and b) for the main system and the auxiliar one.
For larger value of TKEL, the two distributions are different. In fact in this case  due to the opposite kinematical conditions and to the finite geometrical efficiency of the
detector,
different reaction mechanisms are selected on average and the comparison is difficult.
\begin{figure}
\includegraphics[scale=0.43]{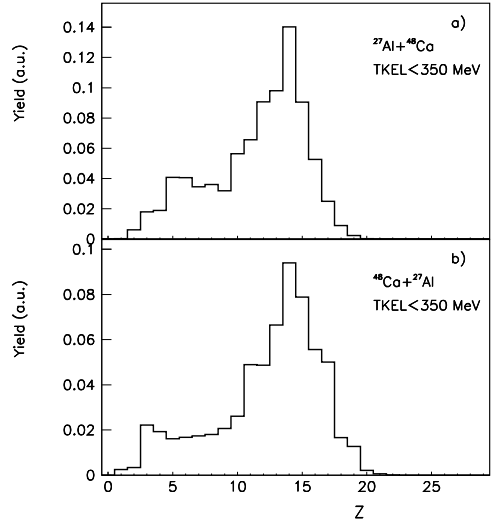}
\vskip -10 pt
\caption{\label{fig2}
For TKEL$<$350 MeV,
(panel a): Charge distribution of PLF's punching through the Si detectors (direct estimate based on the
$\Delta E-E$ method) for the $^{27}Al+^{48}Ca$ system at 40 MeV/nucleon
(panel b): Charge distribution of TLF's stopped in the $Si$ detectors (indirect estimate based on the TOF method) for the $^{48}Ca+^{27}Al$ system at 40 MeV/nucleon}
\end{figure}

The present study  is focused on the investigation of the behavior of the component of $\overrightarrow{D}$
along the beam axis Z,
$D_{Z}$.
As for the case of the TKEL determination and especially  to compensate the eventual systematic errors
in the fragment velocity determination,
instead of
the theoretical value $\overrightarrow{V}_{c.m.}$,
we are necessarily induced to use in eq.(1) the value of
$V_{c.m.,d}^{Z}$  evaluated, event by event, from the velocities along the beam axis
 of all the detected charged particles.
The obtained quantity will be named in the following $D_{Z}^{c}$.
$D_{Z}^{c}$ therefore represents a partial dipolar signal related to the  intrinsic motion
of the subsystem formed by
all the produced charged particle. The contribution related to the global relative motion of the undetected free
 neutrons is not included.
In the next sections we will describe a way to estimate the average total signal associated to $D_{Z}$.
In Fig.3 (panel a)) ,for the main system , we show the correlation plot $Z_{b}$ .v.s. $D_{Z}^{c}$ for the selected events. The ridge in the plot highlights an increasing trend of $\langle D_{Z}^{c} \rangle$ going from negative values to almost zero  for  decreasing  values of $Z_{b}$ respect to $Z_{PLF}$.
\begin{figure}
\includegraphics[scale=0.43]{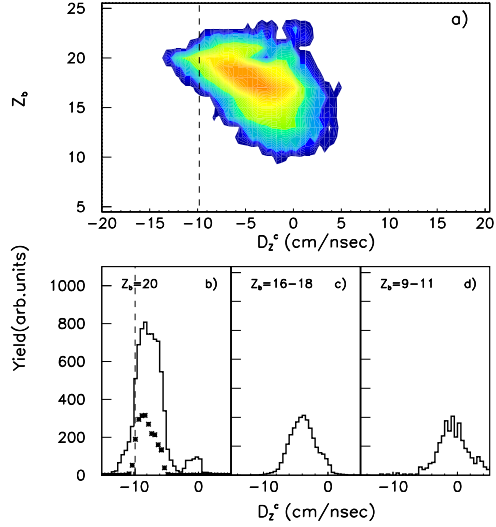}
\vskip -10 pt
\caption{\label{fig3}(Color online)
 (panel a): For the system $^{48}Ca+^{27}Al$ at 40 MeV/nucleon the measured values of  $D_{Z}^{c}$ are plotted for different $Z_{b}$ associated to the selected events (charged multiplicity $m\geq 2$). The dot-dashed vertical lines
indicates the reference limiting values $D_{m}$ (see the text).
(panels b),c),d)): $D_{Z}^{c}$ distributions obtained as projection of the above bi-dimensional plot  for
different $Z_{b}$ intervals. In the panel b) the $D_{Z}^{c}$ spectrum for $Z_{b}$=20 and for quasi-elastic events  TKEL$<$70 MeV is plotted with star symbols.}
\end{figure}
We note that according to the expression given for $D_{m}$, in the initial configuration
the system should exhibit a limiting value of $\langle D_{Z}^{c} \rangle$
(grazing collisions)
close to about -9.8 cm/nsec.
The increasing average values of $\langle D_{Z}^{c} \rangle$ for $Z_{b}$ different
from projectile atomic number (less peripheral collision),
represents a clear signature of the evolution through charge/mass equilibration
values (see next section).
 The limiting value corresponding to almost "grazing" collisions
  (along the beam axis), is evidenced in Fig.3 by dot-dashed
 vertical lines. Panels b),c) and d) show the projections of the bi-dimensional plot for different intervals
 of  $Z_{b}$  and we can clearly see the trend of $\langle D_{Z}^{c}\rangle$.
 In particular, in panel b) the spectrum with star symbols is obtained by imposing a value of TKEL$<$70 MeV
 typical for quasi-elastic processes.
 The large fluctuations of $D_{Z}^{c}$ around the average value are due to
  physical reasons (the particular "history" of the each event) and to the measurement procedure reflecting
  the related uncertainties.
 In Fig.4 we show analogous plots for the reference isospin quasi-symmetric system $^{27}Al+^{40}Ca$ at 40 MeV/nucleon.
 In this case the limiting value for "grazing" collision,  $D_{m}$ is about -2.6 cm/nsec and, how can be clearly
 seen, the experimental plots shows values close to zero and an enhancement  near the $D_{m}$ value (Fig.4b).
 The check on this system as compared to the main one ensure us on the good level of confidence obtained in the determination of
 $\langle D_{Z}^{c} \rangle$.
\section{Comparison with CoMD-II calculations}

To obtain information on the behavior of the Iso-vectorial interaction starting from the present experimental data,
 we have performed CoMD-II \cite{comd,comdii} calculations in the interval of impact parameters $b=4-9$ fm
(the weight of each $b$ is chosen proportionally to $b$  itself).
Dynamical calculations have been followed up to about 500fm/c.
After this primary stage, the produced hot main sources have on average an excitation energy lower than 2.5 MeV/nucelon.
For each generated event, a second stage of a multi-step statistical decay has been simulated  through the
Monte Carlo GEMINI  code \cite{gemini}.
The obtained results have been analyzed with an implemented code that takes into account the main filtering effects of the experimental apparatus.
This include  geometrical acceptance of the identified particles and the main selection criteria used in the analysis of experimental data.
\begin{figure}
\includegraphics[scale=0.43]{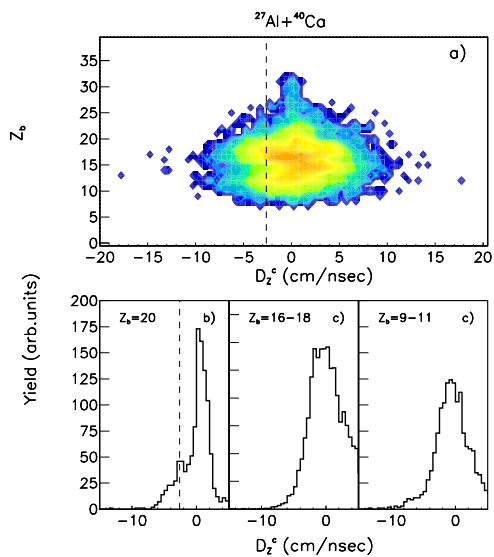}
\vskip -10 pt
\caption{\label{fig4} (Color online) Same as the Fig.2 but for the reference system $^{27}Al+^{40}Ca$}
\end{figure}

\begin{figure}
\includegraphics[scale=0.43]{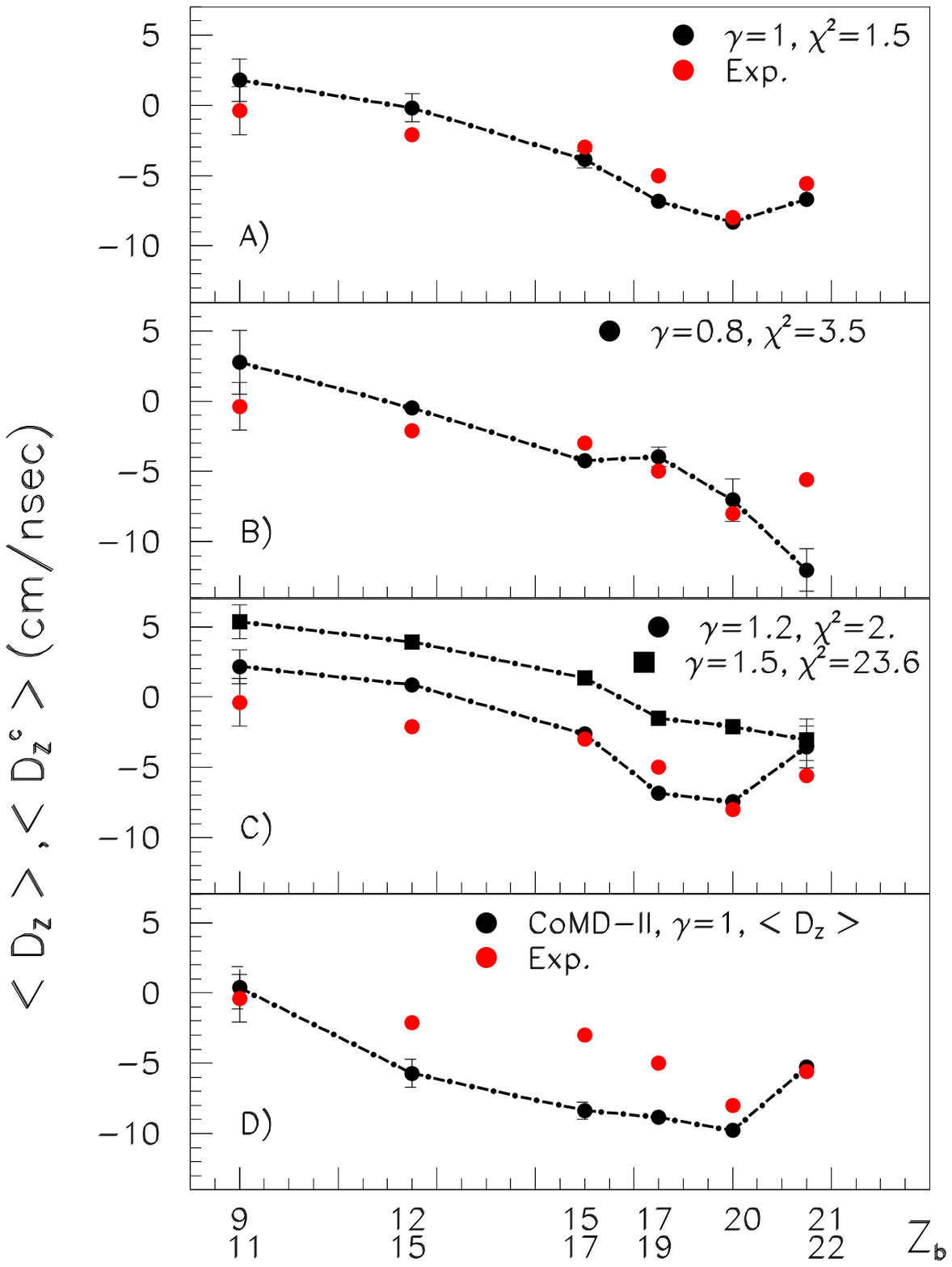}
\vskip -10 pt
\caption{\label{fig5} (Color) For the investigated system $^{48}Ca+^{27}Al$ we plot for different $Z_{b}$ windows
 (the related extremes are indicated by the double labeling in the $Z_{b}$ axis) and for TKEL$<$350 MeV  the measured average dipolar signal $\langle D_{Z}^{c}\rangle$ (red point).  The experimental value are compared in the same figure with the results of COMD-II +GEMINI calculations (black point) filtered through the simulated response of the experimental apparatus and data-analysis selections. Panel A),B),C) show the comparisons for different $\gamma$ values characterizing the iso-vectorial interaction. The $\chi^2$ values with respect to the experimental data are also shown.
 Panel D): the calculated values of $\langle D_{Z}\rangle$ for $\gamma=1$ are compared to the experimental values
 for $\langle D_{Z}^{c}\rangle$.
The error bars represent uncertainties due to the statistics of the simulations and measurement.
The estimated uncertainties  related to experimental values $\langle D_{Z}^{c}\rangle$ are in many cases smaller than the plotted symbols.}
\end{figure}

In particular the CoMD-II calculations have been performed with stiffness parameter values concerning the Iso-vectorial interaction $\gamma=0.5$, $\gamma=0.8$, $\gamma=1$, $\gamma=1.2$, $\gamma=1.5$ and symmetry energy  about $32$ MeV. According to previously performed investigations, the parameters of the effective Skyrme interaction corresponding to
a compressibility of about 220 MeV have been chosen following Ref.\cite{mioskyrme}.
In Fig.5 (panel A),B) C)) we show the comparison of the calculations with the measured value of $\langle D_{Z}^{c}\rangle$ (red points) evaluated starting from the already shown correlation plot in Fig.3. The values are referred to different
$Z_{b}$ windows and  TKEL$<$350 MeV.
The vertical bars indicate the evaluated statistical errors which include also, conservatively, a non-systematic
uncertainty on the measured fragment charge $\Delta Z=\pm 1$.
The corresponding theoretical values are plotted in the different panels with black symbols.
The obtained values for $\gamma=0.5$ are rather different from the experimental results, they are out of the range value plotted in Fig.5. For this case a maximum
value about 7 cm/nsec is obtained for $Z_{b}=9-11$ while the minimum value about -20 cm/nsec is obtained for $Z_{b}=21-22$.
The comparison of the others calculations with the experimental data is  quantified in the figure through the shown $\chi^{2}$ value. According to this parameter, the case with $\gamma=1$ gives the best comparison even if it is
only slightly  better than the case $\gamma=1.2$.
The behavior of  $\langle D_{Z}^{c}\rangle$ (related to only
charged fragments and particles) can be understood in a qualitative way according to the following considerations:
as shown in the previous figure (apart from the case $\gamma=0.8$) for $Z_{b}$ smaller than $Z_{PLF}$
(dominance of mass transfer from PLF to TLF) and for $Z_{b}$ larger than $Z_{PLF}$
(on-set of fusion,incomplete fusion) $\langle D_{Z}^{c}\rangle$  increases
(there are on average more charge/mass symmetric produced fragments). In fact,in this case it can be verified that
$\langle V_{c.m.,d}^{Z} \rangle \leq V_{c.m.}$ therefore
the c.m. of the free neutrons has to be  necessarily larger than $V_{c.m.}$ , that means, on average, more neutron emission from the PLF side.
The case of $\gamma=0.8$ produce instead for  $Z_{b}>Z_{PLF}$ a larger negative value $\langle D_{Z}^{c}\rangle$.
In this case $\langle V_{c.m.,d}^{Z} \rangle \geq V_{c.m.}$ and therefore the cloud
of neutrons has a c.m. velocity smaller than $V_{c.m.}$. In this case we have on average more neutron emission
from the mid-rapidity and TLF side. A closer look to the fragments put in evidence in this case a dominance
of processes producing a PLF fragment, only partially equilibrated in charge/mass and a almost complete disassembly of the TLF.
Up to now we have discussed the behavior of $\langle D_{Z}^{c}\rangle$ which shows a good sensitivity
to the parameters of the effective interaction. To recover information on the global degree
of isospin equilibration
we can evaluate $\langle D_{Z}\rangle$ through calculations by using the same set of parameters
which "best fit" the experimental value of $\langle D_{Z}^{c}\rangle$.
In Fig.5 (panel D)), we compare the calculated values of $\langle D_{Z}\rangle$ for $\gamma=1$ with  CoMD-II+GEMINI (including the efficiency effect) with the experimental values
$\langle D_{Z}^{c}\rangle$.
The connection between the two quantities can be approximated by the following simple
relation: $\langle D_{Z}\rangle \cong \langle D_{Z}^{c}\rangle+Z_{tot}^{d}(\langle V_{c.m.,c}^{Z}\rangle-V_{c.m.})$
where $\langle V_{c.m.,c}^{Z}\rangle$ is the c.m. velocity for the subsystem of the charged particles.
Therefore the second term in the above expression
give us an estimation of the effect associated to the undetected free neutrons \cite{kohley1} which clearly participate in to determining the global degree of isospin equilibration.
Before to conclude this section we add some observations about the "invariance" of the discussed quantities with respect statistical decay modes. As previously observed this was shown in a rather general way in ref.\cite{g1}.
In the present work we have also checked if this insensitivity is still maintained by operating with a realistic efficiency  of the CHIMERA apparatus.
In particular,  in the  "simulated" analysis  we can keep the memory
of the selected primary events from the dynamical model ordered according to the different $Z_{b}$ values
and  TKEL. It is therefore possible to evaluate the same quantity
from the dynamical model $\langle D_{Z}^{D}\rangle$ without taking into account  the
GEMINI secondary decay processes and comparing it to the value of $\langle D_{Z}\rangle$ obtained from the
complete calculations including the efficiency effects. The calculations have been performed for different $Z_{b}$ windows and for
$TKEL<$ 350 MeV.
Table I, as an example, collects the obtained results for different $Z_{b}$ bins by using the calculations for  $\gamma=1$.
\begin{table}[ht]
\caption{For different windows of $Z_{b}$ and  for TKEL$<$ 350 MeV the values (cm/nsec) of $\langle D_{Z}\rangle$
and the corresponding values  of $\langle D_{Z}^{D}\rangle$
are shown for the $^{48}Ca+^{27}Al$ system for $\gamma=1$}
\centering 
\begin{tabular}{c c c} 
\hline 
$Z_{b}$ & $\langle D_{Z}\rangle$(cm/nsec) & $\langle D_{Z}^{D}\rangle$(cm/nsec) \\ [0.6ex]
\hline 
12-15 &-5.73 &-5.9 \\ 
15-17 &-8.36 &-8.34 \\
17-19 &-8.86 &-8.87 \\
20 &-9.79 &-9.78 \\
21-22 &-5.36 &-5.30 \\ [1ex] 
\hline 
\end{tabular}
\label{table:nonlin} 
\end{table}
From the table we can appreciate the close correspondence  between the results after
the GEMINI \cite{gemini} de-excitation stage and
the ones related to the dynamical primary events.
\section{Summary and Outlooks}
In summary, dipolar degrees of freedom for the collision  $^{48}Ca +^{27}Al$ at 40 MeV/nucleon  have been investigated for the first time by means of  4$\pi$ multi-detector CHIMERA. The study has been carried out through the measurement of the observable $\langle D_{Z}^{c} \rangle$ (associated to all the emitted charged particles) and through a related estimation  of the $\langle D_{Z} \rangle$ values
which are closely linked with the global Isospin equilibration process along the beam direction. The discussed insensitivity to the statistical decay
make this observable able to investigate in an exclusive manner the overall dynamics
on this phenomenon.
The study performed on the behavior of $\langle D_{Z}^{c} \rangle$ and $\langle D_{Z} \rangle$
as function of the size of the biggest fragments for TKEL$<350$ MeV shows a noticeable sensitivity of this observable to the parameters of the microscopic effective interaction.
First attempts to reproduce the experimental values of $\langle D_{Z}^{c} \rangle$ in essentially  binary processes
show an overall agrement
with CoMD-II calculations using a stiffness parameter for the iso-vectorial interaction $\gamma\simeq 1\div1.2$.
More experimental investigation
should involve different systems and different reaction mechanisms.
In  particular, a next step forward in this kind of measurements would require a more detailed investigation including a reliable
valuation and/or minimizations of possible systematic errors on the velocity of the produced charged particles.
This would permit a reliable experimental estimation of  $\langle D_{Z} \rangle$ allowing also for a corresponding experimental valuation
of the global effect associated to the dynamically emitted neutrons.
For this purpose long measurements involving targets and projectiles
having the same charge/mass asymmetry (vanishing values of $\langle D_{Z} \rangle$ independently from the reaction mechanism) are necessary.
 A more detailed analysis could also allows to estimate the specific contribution to the equilibration process produced by the prompt and mid-rapidity emission as compared to the transfer of charge/mass between the main partners. The more challenging attempt to investigate on the
other component along the X impact parameter direction
could be of relevant interest. $\langle D_{X} \rangle$ in fact is linked, in a global way, and in an independent way from
statistical emission, with the differential flow of particles
characterized by different charge/mass asymmetries  values \cite{dip}.


\begin{thebibliography}{99}
\bibitem{baorep} Bao-An Li, Lie-Wen Chen, Che Ming Ko, Phys. Rep. \textbf{464}, 113(2008) and reference therin.
\bibitem{barrep} V.Baran, M.Colonna, V.Greco, M. Di Toro, Phys. Rep. \textbf{410}, 335(2005) and reference therin.
\bibitem{bnv} A.Bonasera, F.Gulminelli, and J.Molitoris, Phys. Rep. \textbf{243}, 1(1994).
\bibitem{smm} J.B.Bondorf, A.S.Botvina, A.S. Lljinov, I.N. Mishustin, K.Sneppen, Phys. Rep. \textbf{257}, 133(1995)
\bibitem{piant} S.Piantelli \textrm{et al.}, Phys. Rev. C. \textbf{78}, 064605 (2008).
\bibitem{russ} P.Russotto \textrm{et al.}, Phys. Rev. C. \textbf{81}, 064605 (2010).
\bibitem{g1} M.Papa \textrm{et al.}, Phys. Rev. C \textbf{72},064608(2005) and reference therein (see in particular Appendix D).
\bibitem{dan} P. Danielewicz and J.Lee, Nucl.Phys \textbf{A922} 1(2014).
\bibitem{admard1} G.Ademard \textrm{et al.}, Eur. Phys. Jour. \textbf{A 50} 33(2014).
\bibitem{admard2} G.Ademard \textrm{et al.}, Phys. Rev. C. \textbf{83}, 054619 (2011).
\bibitem{tsang} M.B.Tsang \textrm{et al.},Phys. Rev. Lett. \textbf{92}, 062701 (2004) and reference therein.
\bibitem{marini} P.Marini \textrm{et al.}, Phys. Rev. C. \textbf{87}, 024603 (2013).
\bibitem{kohley} Z.Kohley \textrm{et al.}, Phys. Rev. C. \textbf{85}, 064605 (2012).
\bibitem{lim1} F.Amorini \textrm{et al}, Phys. Rev. Lett. \textbf{102}, 112701(2009).
\bibitem{lim2} G.Cardella \textrm{et al.},Phys. Rev. C. \textbf{85}, 064609 (2012) and reference therein.
\bibitem{natw} Z.Chen \textrm{et al.},Phys. Rev. C. \textbf{81}, 064613 (2010).
\bibitem{yen} K.Brown, S.Hudan, R.T. deSouza, J.Gauthier, R.Roy, D.V.Shetty, G.A.Souliotis,
 S.J.Yennello, Phys. Rev. C. \textbf{87}, 061601 (2013).
\bibitem{levd1} P.Marini, M.F.Rivet, B.Borderie,
\\N.Le Neindre,
A.Chnihi, G.Verde
and J.P.Wieleczko, EPJ Web of Conferences 2, 04003(2010).
\bibitem{napoli}A.Brondi \textrm{et al.},EPJ Web
of Conferences 2, 04002(2010).
\bibitem{isodec}S.Pirrone \textrm{et al.},EPJ Web
of Conferences 17, 16010(2011).
\bibitem{rami} F.Rami \textrm{et al.},Phys. Rev. Lett. \textbf{84}, 1120(2000).
\bibitem{lomb} I.Lombardo \textrm{et al.},Phys. Rev. C. \textbf{82}, 014608 (2010).
\bibitem{sun} Z.Y.Sun  \textrm{et al.},Phys. Rev. C. \textbf{82}, 051603 (2010).
\bibitem{chi1} A.Pagano \textrm{et al.},Nucl.Phys \textbf{A734} 504(2004);A.PAgano, Nucl.Phys.NEws \textbf{22} 25(2012).
\bibitem{rap} G.Giuliani and M.Papa, Phys. Rev. C \textbf{73},031601(2006).
\bibitem{g2} F.Amorini \textrm{et al.}, Phys. Rev C \textbf{58}, 987(1998).
\bibitem{chi2} E.De.Filippo \textrm{et al.}, Phys. Rev. C \textbf{71}, 044602(2005).
\bibitem{alder} M.Alderighi \textrm{et al.}, Nucl. Instr. Meth. \textbf{A489}, 257(2002).
\bibitem{char} R.J.Charity, Phys. Rev C \textbf{58}, 1073(1998) .
\bibitem{comd} M. Papa, T. Maruyama and A. Bonasera, Phys. Rev. C \textbf{64} 024612(2001).
\bibitem{comdii}M.Papa, G.Giuliani A.Bonasera; Jou. of Comp. Physics \textbf{208},403(2005).
\bibitem{gemini} R.J. Charity, D.R. Bowman and Z.H. Liu, R.J. McDonald, M.A. McMahan, G.J. Wozniak,
L.G. Moretto, S. Bradley, W.L. Kehoe, and A.C. Mignerey, Nucl. Phys. A476, 516 (1988).
\bibitem{mioskyrme} M.Papa, Phys. Rev C \textbf{87}, 014001(2013).
\bibitem{kohley1} Z.Kohley \textrm{et al.}, Phys. Rev C \textbf{88}, 041601(2013).
\bibitem{dip} M.Papa M and G.Giuliani, Journal of Physics: Conference Series 312 082034(2011).
\end{thebibliography}
\end{document}